\documentclass[lettersize,journal]{IEEEtran}

\usepackage{doi}

\usepackage{graphicx}
\usepackage{enumerate}
\usepackage{natbib}
\usepackage{url}
\usepackage[misc]{ifsym}
\usepackage{amsmath}
\usepackage{multirow}
\usepackage{cleveref} 
\usepackage{calc}
\usepackage{textcomp}

\begin{document}

 \title{Coexistence assessment and interference mitigation for 5G and Fixed Satellite Stations in C-band in India}

\author{ \href{https://orcid.org/0000-0003-4553-5861}{\includegraphics[scale=0.06]{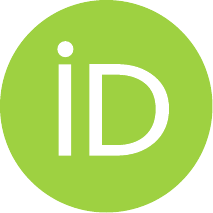}\hspace{1mm}Avinash ~Agarwal} \\
	Telecommunication Engineering Centre\\
 Ministry of Communications\\
 New Delhi, India \\
 \texttt{avinash.70@gov.in} \\
}

\maketitle

\begin{abstract}
In this paper, we present the findings of a study conducted to assess the coexistence of Fifth Generation (5G) wireless networks and Fixed Satellite Station (FSS) receivers in the C-Band (3300-4200 MHz) in India. Through simulations, we evaluate the coexistence feasibility and calculate the minimum separation distances required to mitigate interference, considering factors such as 5G Base Station power, off-axis angle, clutter, filtering, and shielding. Next, we present various interference mitigation techniques, including distance, antenna tilt and height, power control, antenna design, coordination, filtering, and others, aiming for balanced coexistence. The simulation results corroborate the efficacy of these solutions in containing interference from 5G in the C-Band FSS receivers. The paper offers valuable insights into frequency allocation in India and considerations for 5G network design, including site selection and antenna orientation. The insights provided are relevant to other regions facing similar coexistence challenges. Overall, this paper offers a comprehensive overview of 5G and FSS coexistence in the C-band, emphasising the importance of addressing this issue during network design and deployment.
\end{abstract}


\section{Introduction}
\label{intro}
\IEEEPARstart{S}{atellite} communications have become increasingly efficient and affordable due to advancements in encoding, modulation techniques, and the transition from analog to digital technologies. C-band communications have been the backbone of Fixed Satellite Services (FSS), widely used by governments, businesses, and broadcasters for nearly half a century. The robustness of the C-band services and little impact by weather attenuation makes C-band a preferred frequency band for day-to-day business, mission-critical, and disaster recovery. The C-band can provide extensive coverage and is aptly suited for nationwide applications. Due to this reason, the deployment of C-band capacities is economical and reliable in geographies where the terrestrial infrastructure is less developed. In developing nations, the VSAT networks on C-band provide continuous support for services, such as education, telemedicine, and internet telephony. Islands such as Andaman \& Nicobar in India prefer C-band over Ku-band due to its lesser rain attenuation.

Wireless networks based on the International Mobile Telecommunications-2020 (IMT-2020) Standard, commonly known as Fifth Generation (5G), offer significant improvements over existing 4G networks, including very low latency ($\mathrm{\sim}$1 ms), high internet speeds (50 MHz to 10 GHz), and more number of connected devices per unit area. 5G networks are already operational in many countries, with India launching its the 5G services in October 2022. The emergence of 5G, along with advancements in Artificial Intelligence and other technologies, enables various new services such as smart homes, industrial Internet of Things (IIoT), autonomous vehicles, healthcare,  mission-critical applications, and more. Meeting the demands of these 5G-based services requires efficient spectrum allocation and strategic use of available frequency bands.

IEEE \citep{bruder2003ieee} and ITU \citep{iturv431} specify the term C-band for frequencies from 4 GHz to 8 GHz. Historically, satellite receivers have used the C-band from 3.4 to 4.2 GHz (3400-4200 MHz) for downlinking. The FSS Earth Station (ES) equipment, namely the Low Noise Block downconverter (LNB), often operates in the 3400-4200 MHz even though the operating signal might be available only in portions of the band, often above 3600 MHz.

3GPP Release 17, the technical standard for 5G \citep{3gpprelease17}, mentions two frequency bands for 5G New Radio (5G NR), the radio access technology of the 5G mobile networks. Frequency Range 1 (FR1) spans from 410 MHz to 7125 MHz, known as the sub-6 GHz or sub-7 GHz band. In this FR1, bands n77 and n78 are defined in the C-band as 3300-4200 MHz and 3300-3800 MHz, respectively, overlaping with the frequency ranges traditionally used by C-band FSS ES receivers.

Based on Telecom Regulatory Authority of India (TRAI) recommendations \citep{traiauctionrecommendations}, India has allocated 3300-3670 MHz for 5G deployments. There is a guard band of 30 MHz, with the FSS downlinks operating above 3700 MHz. There is, therefore, a possibility of interference with the existing FSS ES receivers from this newly planned use of the band. Today more than 900 registered satellite channels operate in India \citep{broadcastseva}. Allocation of spectrum for 5G up to 3670 Mhz might affect various services, including television services to about 200 million households in India.

\begin{figure}[h]
\centering
\includegraphics[width=\linewidth]{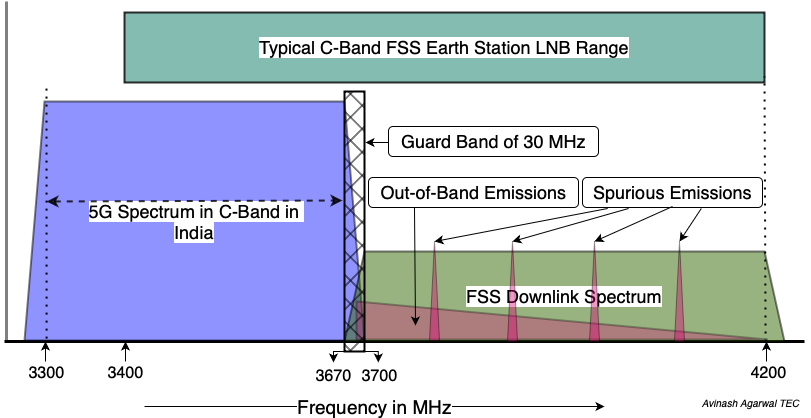}
\caption{Frequency Spectrum in India and various interferences}
\label{fig:1}
\end{figure}

The 5G signals in the C-band fall into the following categories: (1) 5G signals within the allocated spectrum of 3300-3670 MHz, which is the adjacent band for the FSS ES receivers, and (2) unwanted 5G emissions in the 3700-4200 MHz range. These unwanted emissions consist of out-of-band emissions (OOBE) and spurious emissions (SE) as defined by the ITU \citep{3gppts22.261}. Out-of-band emissions are unwanted emissions immediately outside the BS channel bandwidth resulting from the modulation process and non-linearity in the transmitter, excluding spurious emissions. Spurious emissions are emissions caused by unwanted transmitter effects such as harmonics emission, parasitic emission, intermodulation products, and frequency conversion products, excluding out-of-band emissions. Figure \ref{fig:1} shows the interferences in various frequency ranges in India.

The FSS ES LNB amplifier operates linearly only up to a certain power level of the incident signals. In the traditional operating range of the LNB of 3400-4200 MHz, the incident signals could be from the 5G BS, the satellite downlink, or both. As the incident power increases, the amplifier transitions from linear to non-linear behaviour and eventually saturates, resulting in the complete loss of the incoming signals. Satellite signals received by the LNB have much lower power than the higher-powered 5G terrestrial signals, potentially causing overloading/saturation of satellite antenna LNBs and disrupting FSS C-band receivers.

Considering the growing mobile industry and technological advancements, stakeholders prefer the C-band spectrum for 5G applications. Simultaneously, satellite companies have invested significantly in building robust satellite capacities. Balancing the needs of expanding 5G communities while ensuring uninterrupted C-band FSS services is crucial.

This paper presents the results of a coexistence study conducted through simulation for 5G and FSS ES receivers. It further examines the various techniques to mitigate interference from 5G signals to the FSS downlink reception.

\paragraph{Our contributions}
\begin{enumerate}
	\item As far as we know, this study is the first to explore the coexistence of 5G and the FSS BS in the C-band, primarily focusing on the 5G spectrum allocated in India.
	\item The paper also reviews the various techniques to mitigate interference from the 5G BS to the FSS ES reception.
	\item With the rollout of 5G networks in India and other countries, this study will provide valuable assistance in designing networks and selecting optimal 5G BS sites to minimise potential interference to existing FSS ES systems.
\end{enumerate}

\section{Review of coexistence studies}
\label{review}
This section reviews relevant published papers and technical reports on the coexistence of IMT and FSS.

Coexistence studies are available for various regions. For Malaysia, \citep{abdulrazak2008novel} and \citep{abdulrazak2008potential} examine the interference between the FSS ES receiver and IMT-Advanced (4G)/Fixed Wireless Access (FWA) in terms of possible separation distance. These studies include detailed calculations and analysis of path loss, clutter loss, and other factors based on ITU-R recommendations. Their approach is similar to our work, except that our work focuses on interference from 5G and in the Indian environment.

For Brazil, \citep{alexandre2020coexistence} demonstrates the interference problems and image cancellation due to 5G base stations near Satellite Television Receive Only (TVRO) systems in the C-band. It proposes two strategies involving low-cost planar RF filters and increased TVRO LNB 1 dB compression point, validated by ITU-R recommendations, to reduce the required separation distance between 5G base stations (BS) and FSS ES/TVRO users. Similarly, \citep{al2022evaluation} suggests practical solutions, such as installing filters and replacing the satellite receiving LNB to minimise interference from 5G in the FSS ES in Malaysia. The study discusses the ideal distance between antennas and potential protective measures for reducing interference. Our work elaborates on many other interference mitigation techniques in addition to filtering.

The APT Report \citep{apt-112} provides the IMT FSS coexistence case studies in C-Band from China, Indonesia, and Vietnam, exploring frequency subdivisions in the 3100-4200 MHz band. It concludes that terrestrial 4G LTE/5G-NR affects C-band FSS ES receivers, emphasises the importance of mitigation measures, and recommends regulators to analyse spectrum usage and involve stakeholders for protection. Our work focuses on India, assimilating the mitigation techniques discussed in this report and extending them further.

Studies also propose other interference mitigation techniques between 5G and FSS. \citep{hattab2018interference} explores the coexistence of 5G Massive MIMO systems with FSS ES receivers in the 3.7-4.2 GHz band. It analyses aggregate interference and proposes mitigation techniques like protection regions and frequency partitioning. \citep{lagunas20205g} analyses the impact of 5G downlink and uplink on the satellite receiver, considering out-of-band emissions, LNB saturation, and the deployment of Active Antenna Systems (AAS) in terrestrial BSs. It proposes switching off critical emitters and using AAS for beam steering for coexistence. \citep{wei2021distance} presents a distance protection scheme for the Republic of Korea, defining exclusion and restriction zones, and proposes a transmit power control scheme based on Citizens Broadband Radio Service (CBRS). Different scenarios and antenna types are analysed, demonstrating the feasibility of coexistence. The findings provide insights for system parameter selection, aiding in the 5G network design and deployment with satellite systems. \citep{yuniarti2019interference} analyses interference possibilities between 5G and FSS in the extended C-band using deterministic calculations and trial parameters in Indonesia. It concludes that co-channel scenarios require significant separation distance, while adjacent channel scenarios can coexist with a small protection distance using directional antennas. \citep{liu2023guard} proposes a guard band protection method, combined with angular protection and separation, to ensure successful coexistence between 5G BS and FSS ES. GSOA paper \citep{gsoa-adjacent-band} addresses the impact of 5G in-band and out-of-band emission interferences on FSS ES receivers. It demonstrates that filtering should attenuate 5G signals by 60-70 dB to prevent saturation and that even large separation distances may not fully mitigate harmful 5G out-of-band interference. It recommends implementing sufficient guardbands, considering stricter emission profiles for 5G, and registering all FSS earth station receivers for coordination and filter implementation. Expanding on previous studies, we analyse factors impacting separation distance and propose mitigation techniques.

Various technical reports published by standards organisations, industry associations, and corporations also provide insights into different aspects of the coexistence of FWA, 4G, 4.5G, 5G with FSS.

ITU-R Report M.2109 \citep{itur-m2109} suggests separation distances of tens to over a hundred kilometres for deploying IMT-Advanced (4G) systems in FSS bands, reducible by mitigation techniques such as terrain information and site shielding. Adhering to interference levels and coordination agreements is vital for successful coexistence. ITU-R Report S.2199 \citep{itur-s2199} focuses on broadband wireless access (BWA) operating in fixed or mobile modes in the 3400-4200 MHz range, emphasising coordination near FSS earth stations, network parameters' influence on separation distance, and the possibility of mitigation with known station locations. It also concludes that retrofitting FSS with bandpass filters can help but is costly. ITU-R Report S.2368 \citep{itur-s2368} supplements \citep{itur-m2109} and \citep{itur-s2199}, addressing required separation distances for in-band and adjacent-band emissions, LNA/LNB overdrive, and intermodulation interference. Feasible sharing requires specific conditions, but coexistence may be infeasible in typical scenarios, limiting FSS deployments in IMT-Advanced areas. ITU-R Report M.2481-0 \citep{itur-m2481} focuses on enabling the coexistence of IMT and radiolocation services in the 3300-3400 MHz band, offering operational recommendations for urban and suburban areas. Our work differs from these reports as it focuses on IMT-2020 deployment.

ITU-R Report F.2328-0 \citep{itur-f2328} examines the separation distance for IMT and FSS compatibility in the 3400-4200 MHz range through modelling. It concludes that the required separation distance depends on the scenario and environment, ranging from $\mathrm{<}$1 km to nearly 100 km in co-frequency channels. Worst-case adjacent scenarios require $\mathrm{>}$30 km, but practical pointing and mitigation reduce it to a few km with frequency separation. Small cell deployment needs $\mathrm{\sim}$1 km with frequency separation or a few km without it. Our study further explores the factors affecting the minimum separation distance.

\citep{womersley2021review} presents an analysis to address the discrepancies in studies regarding the coexistence of 5G and C-band satellite services. It highlights challenges in co-channel sharing and emphasises the need for accurate modelling in adjacent channel scenarios. It also provides recommended parameters to guide future compatibility studies conducted by administrations.

ITU-R Recommendation M.2101 \citep{itur-m2101} provides a methodology for modelling and simulating IMT networks for sharing and compatibility studies. It encompasses IMT models, system parameters, and simulation steps for analysing emissions and interference, emphasising the importance of realistic modelling and performance comparisons. Our work follows the methodology recommended in \citep{itur-m2101} as far as possible.

Various ITU-R recommendations (\citep{itu-s1432-1}, \citep{itu-sf1486}, \citep{itu-sf1006}, \citep{itu-p452-17}, \citep{itu-s465-6}) provide criteria for calculating interference levels based on multiple factors in different frequency bands. We base our simulation study on these ITU-R recommendations, as discussed in the next section.

\section{Experimental Simulation}
\label{experimentalSimulation}
This section discusses the experimental setup and the assumptions for the simulation study. It also reviews the various ITU-R recommendations and other formulae used for the study. The interference analysis is performed for various separation distances between the 5G Base Station (BS) and the FSS ES receiver, various off-axis angles, different clutter scenarios, and the impact of various factors on the coordination distance or the minimum separation distance between the 5G BS and the FSS ES is analysed.

\subsection{Simulation parameters}
\label{simulationParameters}
\begin{enumerate}
	\item The 5G BS transmitting antenna type is omnidirectional.
	\item The 5G BS EIRP incident towards the FSS ES receiver: various EIRPs considered, default is 72.28 dBm per sector \citep{itu-annex-4.4}.
	\item The 5G BS channel bandwidth is 45 MHz.
	\item The 5G BS antenna height is 10m.
	\item The FSS ES receiver points in azimuth toward the transmitting 5G BS.
	\item The FSS ES receiving antenna is at the boresight of the transmitting antenna.
	\item The FSS ES antenna height is 10m.
	\item The FSS ES antenna elevations: several angles considered between 10\textdegree and 90\textdegree, the default is 10\textdegree.
	\item The FSS ES antenna LNB operating range is 3.4-4.2 GHz.
	\item The propagation mechanism is free space path line-of-sight (LoS).
	\item Percentage of time linked to the propagation model: 20\% time (based on FSS long-term protection criteria time percentage) ITU-R S.1432 \citep{itu-s1432-1}.
	\item Central frequency f is 3.535 GHz (3400 - 3670 MHz).
	\item Clutter category: dense urban, urban, and suburban scenarios considered, default is suburban.
\end{enumerate}

\subsection{5G interference signal level limit}
\label{5GInterferenceSignalLevelLimit}
The maximum permissible interference level is the maximum power of the 5G signals received at the FSS ES receiver beyond which the interference to the FSS receiver becomes significant. It is often defined as a maximum interference over noise level (I/N)$_{5Gmax}$ that should not be exceeded for more than a certain percentage of the time. ITU-R SF.1486 \citep{itu-sf1486} specifies this percentage of time as not more than 20\% of any month. Extrapolating from ITU-R S.1432 \citep{itu-s1432-1}, the long-term protection criteria of I/N not to be exceeded for more than 20\% time of any month, gives:
\begin{equation}
(I/N)_{5Gmax} = -10 dB
\end{equation}

Thus, the level of maximum permissible interference signals I$_{max}$ above the noise floor of the FSS receiver is:
\begin{align}
I_{5Gmax} &= (I/N)_{5Gmax} + N \\
&= (I/N)_{5Gmax} + 10log(kBT)
\end{align}

Considering k (Boltzmann constant) = 1.38 x 10$^{-23}$ J/K, B (Bandwidth 3400-4200 MHz) = 800 MHz, and T (Noise temperature of the receiver) = 100 K \citep{itu-sf1006}, gives:
\begin{equation} \label{eq_i5gMax}
I_{5Gmax} = -129.57 dBW = -99.57 dBm
\end{equation}

\subsection{Strength of the 5G interference signal received at FSS ES}
\label{strengthOfThe5GInterferenceSignalReceivedFSS}
ITU-R SF.1486 \citep{itu-sf1486} provides the relations of the total interference power I, from an FWA TS at the VSAT input to the antenna as a function of the separation distance, d, and the off-axis angles of the VSAT ($\mathrm{\phi}$) and FWA TS ($\alpha$) for various assumptions. Adapting the same for 5G BS, we have:
\begin{equation} \label{eq_stength5GInterference}
I_{5G} = EIRP_{5G}(\alpha) - L_{5G}(d) + G_{FSS}(\mathrm{\phi}) - R - F
\end{equation}

Where EIRP$_{5G}$($\alpha$) is the off-axis Effective Isotropic Radiated Power from the 5G BS transmitter, L$_{5G}$(d) is the path loss between the FSS ES and the 5G BS antennas, G$_{FSS}$($\mathrm{\phi}$) is the FSS ES off-axis antenna receive gain in the direction of the interfering 5G BS transmitter, R is the isolation from the site shielding, F is the centre frequency offset factor. The impact of each of these factors is examined as follows.

\subsubsection{5G BS EIRP}
\label{5GBS}
Effective Isotropic Radiated Power (EIRP) from the 5G BS transmitter toward the FSS ES receiver depends on the power radiated by the transmitter, its gain, and the angle of the boresight of the transmitter with the ES receiver. In our simulation, we assume the 5G transmitting antenna to be omnidirectional and the receiving antenna to be at the boresight of the transmitting antenna.

The EIRP may vary depending on the deployment scenario such as rural, suburban, or urban, and also on the type of BS such as macro, micro, or small cell. ITU-R \citep{itu-annex-4.4} specifies the maximum base station output power/ sector (EIRP) for rural/ suburban/ urban macro BS as 72.28 dBm and urban small cell (outdoor)/ microcell as 61.53 dBm. Accordingly, in our simulation, we have considered the 5G BS EIRP incident towards the FSS ES receiver as 72.28 dBm per sector. We also calculate the minimum separation distance required between the two for BS EIRP values of 42, 52, 62, and 72 dBm.

\subsubsection{Multiple 5G BS scenario}
\label{multiple5GBS}
We consider two scenarios.

Case 1: In the first case, we consider a single 5G BS carrier of 45 MHz. The maximum EIRP towards the FSS ES receiver is 72.28 dBm as mentioned above.

Case 2: In the worst case, we consider multiple 5G BS carriers covering the entire 270 MHz overlap (3400-3670 MHz), with each carrier having a bandwidth of 45 MHz and maximum EIRP towards the FSS ES as 72.28 dBm. Accordingly, the total maximum EIRP falling within the LNB receiver for all the carriers together is:
\begin{multline}
EIRP_{WorstCase} = 72.28 + 10log10 (270/45) \\= 80.06 dBm
\end{multline}

This is about 7.78 dBm more than that of case 1.

\subsubsection{Propagation losses}
\label{propagationLoses}
The 5G interfering signal undergoes losses as it propagates from the BS to the FSS ES. ITU-R P.452-17 \citep{itu-p452-17} gives various long-term as well as short-term interference propagation mechanisms including line-of-sight (LoS), diffraction, tropospheric scatter, surface ducting, and clutter. In our simulation, we consider the free space path (LoS) and also ignore the attenuation due to atmospheric gases for simplicity and without loss of generality.

The loss incurred by the 5G signal (L$_{fsp}$) as it propagates in a straight line through a vacuum with no absorption or reflection of energy from nearby objects depends on frequency (f) or wavelength ($\lambda$) distance (d):
\begin{equation}
(L_{fsp}) = (4{\pi}d/\lambda)^{2} = (4{\pi}fd/c)^{2}
\end{equation}

In dB,
\begin{multline}
L_{fsp} = 10log(4{\pi}fd/c)^{2} \\= 20log(4\pi/c) + 20log(f) + 20log(d)
\end{multline}

Considering c = 3x10$^{8}$ m/s, f in GHz and d in Km:
\begin{equation} \label{eq_propagationLoss}
L_{fsp} = 92.44 + 20log(f_{GHz}) + 20log(d_{Km})
\end{equation}

In our simulation, we consider f = 3.535 GHz as the central frequency of the 3400 - 3670 MHz range.

Considering the attenuation due to atmospheric gaseous absorption (A$_{g}$) and due to clutter (A$_{h}$), the equation becomes:
\begin{multline} \label{eq_propagationLossWithAttenuation}
L_{fsp} = 92.44 + 20log(f_{GHz}) \\+ 20log(d_{Km}) + A_{g} + A_{h}
\end{multline}

This conforms with ITU-R P.452-17 \citep{itu-p452-17} and ITU-R S.1432 \citep{itu-s1432-1}.

\subsubsection{Off-axis angle and antenna gain}
\label{offAxisAngle}
ITU-R P.452-17 \citep{itu-p452-17} gives the off-axis/ off-boresight angle, $\varphi$ i.e., the angle between the interference axis and direction of the main lobe (ES antenna orientation) of the FSS ES as:
\begin{equation}
\varphi = arccos(cos(\alpha)cos(\epsilon)cos(\mathrm{\vartheta}) + sin(\alpha)sin(\epsilon))
\end{equation}

where, $\alpha$ = ES elevation angle, $\epsilon$ = (h$_{ES}$ $\mathrm{-}$ h$_{BS}$)/d $\mathrm{-}$ d/(2r), $\mathrm{\vartheta}$ = azimuth angle of the BS w.r.t. the ES main lobe, and r = effective Earth radius = 8.5 $\mathrm{\times}$ 10$^{6}$ m.

We assume the FSS ES receiver to be pointing in azimuth towards the transmitting 5G BS. As mentioned above, we also assume the 5G transmitting antenna to be omnidirectional and the receiving antenna to be at the boresight of the transmitting antenna. So, in our case $\mathrm{\vartheta}$ = 0$^{\circ}$ and $\varphi$ $\approx$ $\alpha$.

ITU-R S.465-6 \citep{itu-s465-6} provides the variation of FSS ES receiver antenna gain with the off-boresight angle for frequencies in the range from 2 to 31 GHz as:
\begin{align} \label{eq_gainFSS}
G_{FSS}(\varphi) &= 32 - 25log\varphi \ dBi \ for \ \varphi_{min} \mathrm{\le} \varphi \mathrm{<} 48\mathrm{^\circ}\\
&= -10 \ dBi \ for \ 48\mathrm{^\circ} \mathrm{\le} \varphi \mathrm{\le} 180\mathrm{^\circ}
\end{align}

where,
\begin{align*}
\varphi_{min} &= 1\mathrm{^\circ} \ or \ 100 \lambda/D \ degrees, \\ &whichever \ is \ the \ greater, \ for \ D/\lambda \mathrm{\ge} 50.\\
&= 2\mathrm{^\circ} \ or \ 114 (D/\lambda)^{-1.09} \ degrees, \\ &whichever \ is \ the \ greater, \ for \ D/\lambda \mathrm{<} 50.
\end{align*}

where D is the diameter of the disc antenna.

At 3.5 GHz, $\varphi$ $\mathrm{>}$ 4.8$\mathrm{^\circ}$ for D = 1.8 m and $\varphi$ $\mathrm{>}$ 3.6$\mathrm{^\circ}$ for D = 2.4 m \citep{itu-sf1486}.

We evaluate the coordination distance and the interference signal power for various angles between 10$\mathrm{^\circ}$ and 90.$\mathrm{^\circ}$

\subsubsection{Clutter Losses}
\label{clutterLosses}
ITU-R P.452-17 \citep{itu-p452-17} provides the expression for the additional loss due to local clutter as:
\begin{multline}
A_{h} = 10.25F_{fc}.e^{-d_{k}} (1 - tanh(6*(h/h_{a} - 0.625))\\ - 0.33 dB
\end{multline}

where, F$_{fc}$ = (0.25 + 0.375*(1+tanh(7.5*(f $\mathrm{-}$ 0.5)))) = 1, d$_{k}$ = distance (km) from the nominal clutter point to the antenna, h = antenna height (m) above local ground level, h$_{a}$ = nominal clutter height (m) above local ground level.

\citep{itu-p452-17} also defines various clutter categories and provides their reference nominal clutter height and distance from the antenna, as tabulated in table \ref{table:1}.

\begin{table}[h]
\small
\centering
\begin{tabular}{|c|p{25mm}|p{25mm}|}
\hline
Clutter category & Nominal clutter height h$_{a}$ (m)     & Nominal distance dk (km) \\ \hline
Village centre & 5 & 0.07 \\ \hline
Suburban & 9 & 0.025 \\ \hline
Dense suburban & 12 & 0.02 \\ \hline
Urban & 20 & 0.02 \\ \hline
Dense urban & 25 & 0.02 \\ \hline
High-rise urban & 35 & 0.02 \\ \hline
Industrial zone & 20 & 0.05 \\ \hline
\end{tabular}%
\caption{Nominal clutter heights and distances (ITU-R P.452-17 \citep{itu-p452-17})}
\label{table:1}
\end{table}

We calculate the clutter losses for the different categories and evaluate their impact on the coordination distance. In our simulation, the default is the suburban category with the FSS ES antenna height of 10 m.

\subsubsection{Bandpass filter and shielding}
\label{bandpassFilterAndShielding}
We examine the impact of shielding and using a bandpass filter by varying the values of R in equation \ref{eq_stength5GInterference} above.

\subsection{Strength of the C-Band satellite signal received at FSS ES}
\label{strengthOfCBandSatellite}
The potential satellite signal level in 3700-4200 MHz is estimated as follows:
\begin{equation} \label{eq_cSat}
C_{Sat} = EIRP_{Sat} + 10log(N_{Carrier}) - L_{Sat} +G_{rx}
\end{equation}

Assuming the free space propagation mechanism and using equation \ref{eq_propagationLoss}, we get the path loss for a distance of 42,000 Km from the satellite to the FSS ES as:
\begin{align} \label{eq_satLoss}
L_{Sat} &= 92.44 + 20log(3.95) + 20log(42000)\\
&= 196.897 dBW
\end{align}

Assuming 13 carriers of 36 MHz each, transponder EIRP of 40 dBW per transponder, and the FSS ES receiver antenna gain of 43 dB, the strength of the C-Band satellite signal received at FSS ES is obtained by substituting values in equation \ref{eq_cSat}:
\begin{align}
C_{Sat} &= 40 + 10log(13) - 196.897 + 43\\
&= -102.76 dBW = -72.76 dBm
\end{align}

\subsection{The total strength of the signals received at the FSS ES}
\label{totalStrengthOfSignalReceivedFSS}
The total power received by the FSS ES antenna I$_{tot}$ in its operating frequency band is the sum of the total power levels received from all the 5G BS as per equation \ref{eq_stength5GInterference} and the total power of all the carriers from the satellite as given by equation \ref{eq_cSat}.
\begin{equation} \label{eq_totalStrength}
I_{tot} = 10log\mathrm{\{}10^{\mathrm{\wedge}}(I_{5G}/10) + 10^{\mathrm{\wedge}}(C_{Sat}/10)\mathrm{\}}
\end{equation}

\subsection{LNB non-linearity and saturation}
\label{LNBSaturation}
The low-noise block down-convertor (LNB) of a satellite earth station consists of an amplifier, a down-convertor, and then another amplifier. The first-stage amplifier amplifies the received signals while adding very less noise. The down-convertor then converts the signal to intermediate frequencies (IF). The second amplifier amplifies the IF signal and feeds it to the modem receiver. The amplifiers should operate in the linear region where the gain is fixed and a change in the input signal produces a proportional change in the output signal. However, the amplifiers have a maximum output power limit and if the input signal power is increased beyond a limit, then the amplifier enters the non-linear or the gain compression stage, where the gain decreases and the output is no longer in the same ratio as the input signal. On further increasing the input power, the amplifier reaches the maximum output power level, where any change in the input signal does not produce any change in the output. This is referred to as the saturation stage and the input signals are completely lost.

\begin{figure}[h]
\centering
\includegraphics[width=\linewidth]{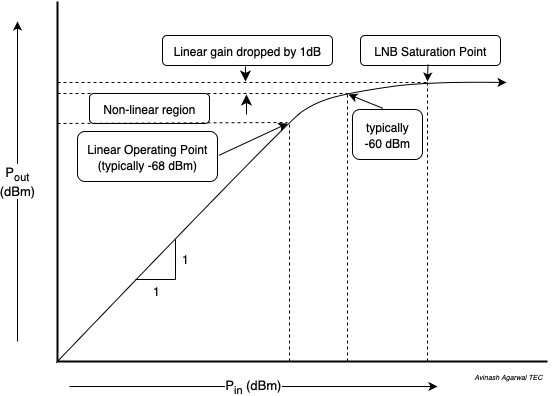}
\caption{LNB input-output power curve}
\label{fig:2}
\end{figure}

Studies show that the typical LNBs operate in the linear region for the input signal strengths less than -68 dBm, followed by the non-linear region from -68 dBm to -60 dBm, and the saturation region beyond -60 dBm. This is represented in figure \ref{fig:2}.

The total power received by the LNB from both the satellite transponders (wanted) and the 5G BS (unwanted), as per equation \ref{eq_totalStrength} above, should therefore be typically less than -68 dBm.

\subsection{Minimum separation distance}
\label{minimumSeparationDistance}
By substituting the formula for L$_{fsp}$ from equation \ref{eq_propagationLossWithAttenuation} in equation \ref{eq_stength5GInterference}, and rearranging, we get:

\begin{multline} \label{eq_minSeparationDist}
20log(d_{Km}) = EIRP_{5G}(\alpha) - I_{5G} - 92.44 \\- 20log(f_{GHz}) - A_{g} - A_{h} + \\G_{FSS}(\varphi) - R - F
\end{multline}

The minimum separation distance between the 5G BS and the FSS ES receiver is obtained when the maximum permissible value of I$_{5G}$ is put in equation \ref{eq_minSeparationDist}. When the 5G BS signals are in the same frequencies as that of the FSS ES, as in the case of Out of Band emissions and spurious emissions, then lesser of the threshold limit as per equation \ref{eq_i5gMax} above and the LNB linear-region limits should be considered. When the 5G signals are in the adjacent band of 3300-3670 MHz and do not overlap with the FSS downlink signals in the 3700-4200 MHz range, then the LNB linear-region limits should be considered.

We examine the various minimum separation distances required for different values of 5G EIRP, off-axis angles, clutter scenarios, filtering, shielding, etc.

\section{Results}
\label{results}
The observations are analysed to examine the impact of distance, 5G BS radiated power, multiple BS, off-axis angle of the FSS ES, filtering, clutter, and shielding for mitigating the 5G interference.

\subsection{Signals received by the FSS ES receiver}
\label{signalsReceivedFSS}
The total power received by the FSS ES is the total of the signals received from various 5G base stations and the satellite signals in the operating range of the FSS receiver antenna (3.4-4.2 GHz).

The 5G signals in the allocated band of 3300-3670 MHz do not overlap with the FSS downlink signals in the 3700-4200 MHz band, still they can interfere with the FSS signal reception due to their high power levels. 5G signals are much stronger and can saturate the FSS ES LNBs.

The Out of Band Emissions (OOBE) and Spurious Emissions (SE) from the 5G base stations in the 3700-4200 MHz range can directly interfere with the FSS downlink signals as ES receivers as the frequency ranges are the same. They can also saturate the FSS ES LNBs if their power levels are high, leading to a complete loss of signals. Bandpass filters installed at LNBs cannot filter out these unwanted emissions because they fall in the pass band of these filters. 3GPP Standards specify sharp spectrum mask requirements for 5G BS transmitters. Spectrum masks are typically defined in terms of a maximum allowed power level, expressed in dBm, for a given frequency range. 5G equipment complying with these standards will limit the power of these OOBE and SE signals.

Low-power private captive networks proposed in the 3700-3800 MHz range could also interfere with the C-band FSS ES receivers in their vicinity.

\subsection{Interference (5G) signal power versus distance}
\label{interference5GSignalPowerVsDistance}
\begin{figure}[h]
\centering
\includegraphics[width=\linewidth]{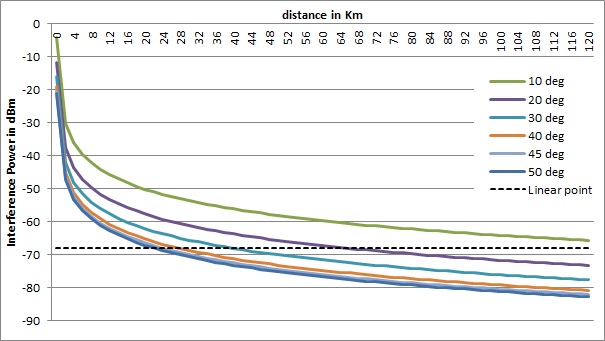}
\caption{Interference (5G) signal power versus distance}
\label{fig:3}
\end{figure}

Figure \ref{fig:3} plots the power of the 5G interference signals as a function of the separation distance between the 5G base station and the FSS ES for various off-axis angles. The graphs are for the suburban scenario without any filtering or shielding. As calculated in the previous section, the maximum 5G signal at the FSS ES that may not cause interference is -99.57 dBm. However, since the 5G and the FSS receivers would have separate frequency bands, only the LNB saturation is relevant in this case. The LNB saturation points and LNB linear operations points for various off-axis angles are obtained for the power levels of -60 dB and -68 dB respectively. For LNB to operate in the linear region, the separation distance should be more than that at the -68 dB point.

\subsection{Impact of 5G BS power}
\label{impactOf5gBSPower}
\begin{figure}[h]
\centering
\includegraphics[width=\linewidth]{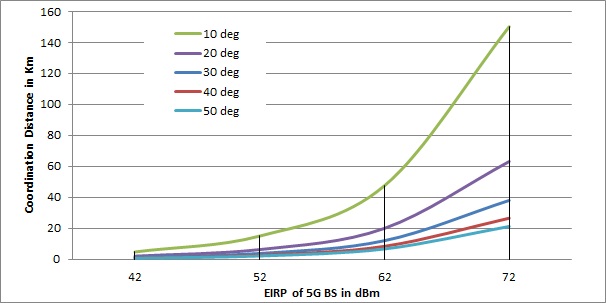}
\caption{Impact of 5G BS power}
\label{fig:4}
\end{figure}

Figure \ref{fig:4} shows the coordination distances or the minimum separation distances between the 5G base station and the FSS ES receiver as a function of the total EIRP falling within the LNB receiver to avoid LNB non-linearity for various off-axis angles. As the power of the 5G signal increases, the minimum separation distance required also increases.

\subsection{Interference from multiple 5G BS}
\label{interferenceFromMultiple5gBS}
\begin{figure}[h]
\centering
\includegraphics[width=\linewidth]{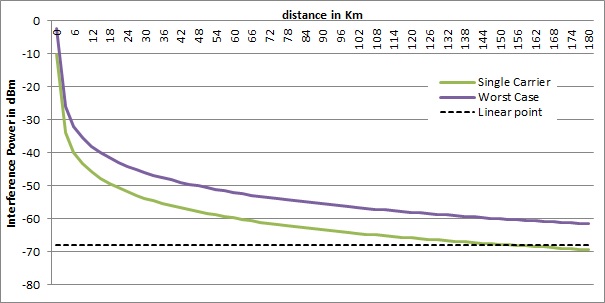}
\caption{Interference from multiple 5G BS}
\label{fig:5}
\end{figure}

Figure \ref{fig:5} plots the power of the 5G interference signal as a function of the separation distance for the two scenarios mentioned in para III.C.(2). The worst-case scenario shows the interference from multiple 5G carriers covering the entire 270 MHz overlap (3400-3670 MHz). The total EIRP incident on the FSS ES receiver is nearly 7.78 dBm higher than in the case of a single carrier. The coordination distances for the single-carrier and the worst-case scenarios are 155 Km and 380 Km respectively.

In practical situations, the worst-case scenario is unlikely as there is a very low possibility of multiple 5G carriers covering the entire 270 MHz frequency range and radiating in alignment with the boresight of the FSS BS antenna.

\subsection{Impact of using a filter}
\label{impactOfUsingFilter}
\begin{figure}[h]
\centering
\includegraphics[width=\linewidth]{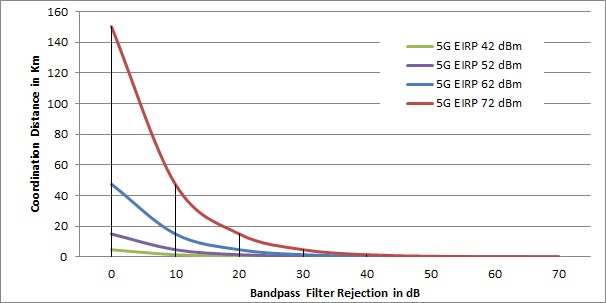}
\caption{Impact of using a filter}
\label{fig:6}
\end{figure}

\begin{figure}[h]
\centering
\includegraphics[width=\linewidth]{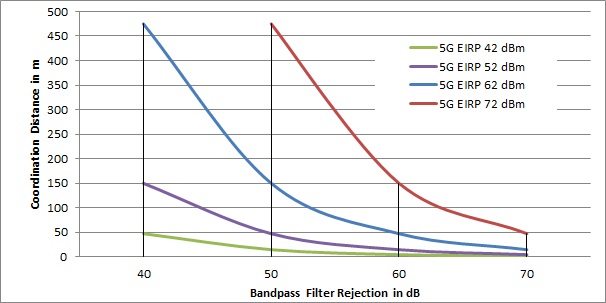}
\caption{Impact of using a filter (enlarged)}
\label{fig:7}
\end{figure}

Figures \ref{fig:6} and \ref{fig:7} reveal that a bandpass filter considerably reduces the coordination distances for the various scenarios. For the suburban clutter scenario with a 10$^{\circ}$ off-axis angle and 72 dBm EIRP of the 5G BS considered in the experiment, using a filter of 60 dB reduces the coordination distance to just 150 m compared to 150 Km when not using a filter. The corresponding figure is 1.5 Km when using a filter of 40 dB.

The minimum separation distance required for a given site acts as the parameter for designing or selecting the appropriate filter.

\subsection{Impact of off-axis angle}
\label{impactOfOffAxisAngle}
\begin{figure}[h]
\centering
\includegraphics[width=\linewidth]{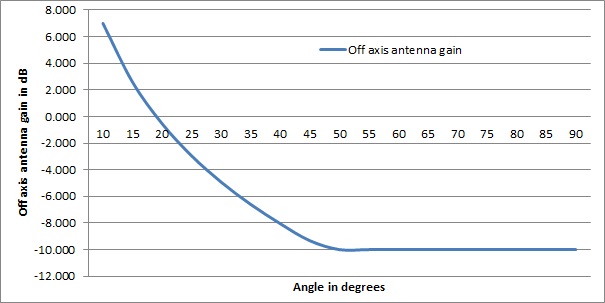}
\caption{FSS ES off-axis antenna gain}
\label{fig:8}
\end{figure}

\begin{figure}[h]
\centering
\includegraphics[width=\linewidth]{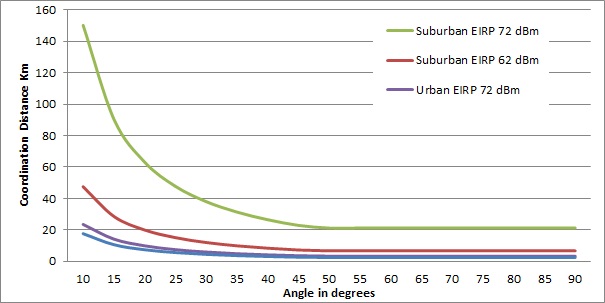}
\caption{coordination distances for various off-axis angles}
\label{fig:9}
\end{figure}

Increasing the off-axis angle up to 48$^{\circ}$ decreases the FSS ES off-axis antenna gain in the direction of the interfering 5G BS, as shown in figure \ref{fig:8}. The coordination distance also decreases accordingly. Figure \ref{fig:9} plots the coordination distances for various off-axis angles according to ITU-R S.465-6 \citep{itu-s465-6} as per equation \ref{eq_gainFSS} for suburban, urban, and dense urban scenarios. For angles more than 48$^{\circ}$, the coordination distance remains the same.

For a suburban clutter scenario with 5G BS EIRP of 72 dB without filter, the coordination distance at an elevation of 10$^{\circ}$ is 150 Km, which reduces to 21 Km at angles greater than 48$^{\circ}$. Also, the ES antenna gain towards the 5G interfering BS for a given elevation of more than 10$^{\circ}$ is independent of the ES antenna size (either 32 - 25log(off-axis) or -10 dBi).

If the angle is less than 10$^{\circ}$ or the 5G base station transmits directly into the bore sight of the FSS receiver, then even filters would not be effective in mitigating interference.

\subsection{Clutter scenarios}
\label{clutterScenarios}
Figure \ref{fig:9} shows the coordination distance for suburban, urban, and dense urban clutter scenarios as per ITU-R P.452-17 \citep{itu-p452-17}. The nominal clutter height for a suburban propagation environment specified in \citep{itu-p452-17} is 9m. In our simulation, both antennas are above this level, so there is no clutter loss in the suburban case. For urban and dense urban scenarios, the clutter losses are around 16.1 dB and 18.6 dB respectively, and accordingly, the coordination distances are lesser compared to the suburban case.

\subsection{Impact of shielding}
\label{impactOfShielding}
Proper shielding of the FSS ES receive antenna has a similar impact as that of a filter. Shielding allows using a filter with lesser attenuation for a similar reduction in the coordination distances. ITU-R SF.1486 \citep{itu-sf1486} mentions that an exposed roof-top mounting for a VSAT protected by suitable shielding is more desirable than siting at ground level and that the effect of shielding can be approximately 33 dB, as calculated per Recommendation ITU-R P.526. FSS ES should be sited to maximise shielding from possible 5G interferences. They should not be installed higher than is necessary for the application. Available terrains, such as depressions, tree grooves, natural screening, and artificial obstructions such as buildings, should be judiciously used to increase obstructive/ diffraction losses of the interfering signals.

\section{Discussion and interference mitigation techniques}
\label{discussion}
The results show that the interference level depends upon several factors, such as the separation distance, 5G BS transmit power, antenna gain, antenna height, the direction of the antenna's main beam, the off-axis angles, terrain, atmospheric conditions, etc. We discuss the various techniques that can mitigate interference between the 5G NR systems and C-band fixed satellite services receivers:

\begin{enumerate}
\item Network planning: Careful network planning helps to minimise the risk of interference between 5G NR and C-band fixed satellite services. This involves optimising the location and orientation of antennas, as well as ensuring adequate separation between the two systems in the frequency domain.
\item Filtering: Properly designed bandpass filters block adjacent band interference and allow only the desired signal to pass through to the receiver. They effectively block the higher power 5G NR signals from saturating the LNBs of the FSS receivers and considerably reduce the minimum separation distance.
\item Shielding: Shielding techniques, such as the use of electromagnetic shielding or mechanical isolation, can also reduce the risk of interference between 5G NR and C-band FSS systems. This is achieved by selecting site locations that are shielded from 5G NR signals or by using shielding materials to block the transmission of 5G NR signals.
\item Electromagnetic shielding: Electromagnetic shielding involves the use of shielding enclosures or shielded cables to isolate the FSS receiver from 5G NR signals.
\item Mechanical isolation: Mechanical isolation involves physically separating the 5G NR and FSS systems or the use of barriers, such as walls, buildings, or terrain features, to attenuate the 5G NR signals.
\item Distance: One approach to mitigating interference is to ensure that the site locations selected for 5G NR and C-band FSS systems are sufficiently far apart.
\item Antenna height: The FSS earth station antennas generally face upwards to receive signals from satellites up in the sky, so interference can be reduced by installing the 5G BS antenna at a height lower than the FSS receiver antenna. However, lowering the 5G macro cell beyond a certain height may not be possible as it may degrade the cell coverage.
\item Antenna tilt: Vertical down tilting of the 5G BS antenna can significantly reduce the interference with the FSS earth station.
\item Orientation: The orientation of the antennas used by 5G NR and C-band FSS systems can also be used to mitigate interference. For example, by selecting site locations where the antennas are oriented in opposite directions, the risk of interference can be reduced.
\item Microcells: Another technique is to deploy microcells instead of macrocells in the vicinity of FSS ES receivers. These have lower power than the macrocells and can be installed at lower heights, thus reducing interference significantly.
\item Directional antennas: The 5G NR system can use directional antennas that are oriented away from the direction of the FSS receiver, thus limiting the 5G NR signal power transmitted in the direction of the FSS receiver.
\item MIMO, adaptive beamforming, and null steering: The 5G NR system can use MIMO (multiple-input multiple-output) to null the radiation pattern in the direction of FSS earth stations to reduce the risk of interference. This is accomplished by using multiple antennas and carefully controlling the phase and amplitude of the signals from each antenna in the array to create a focused beam of energy towards the intended receiver (called beamforming) and to create a null in the direction of the FSS receiver (called null steering).
\item Power limiting: By reducing the transmit power, the 5G NR systems can operate at a lower level, which can help to minimise interference to the FSS receivers.
\item Power control: Techniques such as dynamic power control, which adjusts the transmit power of the 5G NR system based on the distance to the receiver and the channel conditions, or power backoff, reduce the transmit power of the 5G NR system when it is operating close to the C-band fixed satellite services receiver.
\item Spectrum masking: By adhering to the spectrum mask specified in the 3GPP technical specifications, it is possible to minimise the risk of interference due to the out-of-band (OOB) and spurious emissions (SE) from the 5G base stations with the FSS receivers that operate in the same frequency range. Bandpass filters cannot filter out these signals.
\item Coordination and sharing: Coordination and sharing agreements between 5G NR systems and C-band fixed satellite services can also be used to minimise interference. These agreements can specify the frequency bands and geographic areas in which the two types of systems can operate, as well as any other technical or operational requirements that may be necessary to ensure that the systems can coexist effectively.
\item Cooperation: Cooperation between 5G NR and C-band fixed satellite services can help to ensure that the two types of systems can coexist effectively. This may involve the use of protocols or standards that allow the two types of systems to communicate and coordinate with each other, such as the use of signalling protocols to exchange information about the operating conditions of the two systems.
\item Dynamic frequency selection: Dynamic frequency selection (DFS) allows 5G NR systems to automatically detect and avoid frequencies being used by C-band fixed satellite services.
\item Diversity techniques: Diversity techniques, such as spatial diversity or frequency diversity, improve the performance of the satellite receiver in the presence of interference. Spatial diversity involves the use of multiple antennas at the satellite receiver to improve the signal-to-noise ratio, while frequency diversity involves the use of multiple frequencies to reduce the impact of interference on the signal.
\end{enumerate}

The effectiveness of these techniques will depend on various factors, such as the specific operating conditions and the level of interference that is present. It may also be necessary to consider other factors, such as the cost and feasibility of implementing these solutions. The most effective approach to mitigating interference may require a combination of these techniques.

\section{Conclusion and future work}
\label{conclusion}
The paper explores the coexistence of 5G wireless networks and FSS receivers in the C-Band, specifically focusing on the spectrum allocated in India for 5G services. By employing simulations, the paper assesses the feasibility of coexistence and computes the minimum separation distances necessary to mitigate interference. Our analysis considers several critical factors, including 5G Base Station power, off-axis angle, multiple 5G BS, clutter, filtering, and shielding.

Furthermore, based on the simulation results, the paper presents a range of interference mitigation techniques, encompassing shielding, distance management, antenna tilt and height optimization, power control, antenna design, and coordination strategies. These techniques help in achieving balanced coexistence between 5G and FSS systems. 

The implications of our findings extend beyond India as they apply to other regions facing similar coexistence challenges between 5G and FSS systems. The comprehensive overview presented in this paper contributes to the body of knowledge on the coexistence of 5G NR and FSS in the C-Band and offers valuable guidance for policymakers, regulators, and industry stakeholders.

As the deployment of 5G networks and FSS systems continues to expand, it is crucial to consider the coexistence issue during the design and deployment phases. Telecom operators are currently rolling out 5G services in India and other countries. This study will assist both telecom operators and FSS ES operators in planning their networks to mitigate interference from 5G signals.

\textbf{Future Work:} Future work can focus on:

Experimental Validation: Conduct field tests to validate simulation results, measure interference levels, signal quality, and coexistence parameters in various locations, and evaluate the real-world performance of proposed interference mitigation techniques.

Economic Analysis: Assess the cost-effectiveness of implementing interference mitigation techniques and coexistence solutions. Evaluate expenses associated with filter deployment, antenna upgrades, infrastructure modifications, etc. Compare costs with potential benefits of improved coexistence.

\bibliographystyle{unsrtnat}

\end{document}